\documentstyle[prb,aps]{revtex}
\draft
\begin{document}
\title{Phonon-drag effects on thermoelectric power}
\author{M. W. Wu}
\address{
Department of Physics, University of Science and Technology of China,
96 Jingzhai Road, Hefei, Anhui, 230026, China}
\address{Department of Physics and Engineering Physics, Stevens Institute
of Technology, Hoboken, NJ 07030}
\author{N.J.M. Horing and H.L. Cui}
\address{Department of Physics and Engineering Physics, Stevens Institute
of Technology, Hoboken, NJ 07030}
\maketitle
\begin{abstract}
We carry out a calculation of the phonon-drag contribution $S_g$
to the thermoelectric
power of bulk semiconductors and quantum well structures for the first time
using the balance equation transport theory extended to the weakly
nonuniform systems. Introducing wavevector and
phonon-mode dependent relaxation times due to phonon-phonon
interactions, the formula obtained can be used not only at
low temperatures where the phonon mean free path is determined by
boundary scattering, but also at high temperatures.
In the linear transport limit, $S_g$ is equivalent to the
result obtained from the Boltzmann equation with a relaxation time
approximation. The theory is applied to experiments and agreement is
found between the theoretical predictions and experimental results.
The role of hot-electron effects in $S_g$ is discussed.
The importance of the contribution of $S_g$ to thermoelectric power
in the hot-electron transport condition is emphasized.
\end{abstract}
\bigskip


\section{Introduction}

Recently the study of thermoelectric power has attracted intensified
interest both theoretically and experimentally in various systems:
mesoscopic quantum dots \cite{beenakker,molenkamp}, quantum
wires \cite{kearney}, heterojunctions and quantum
well structures
\cite{nicholas,fletcher1,fletcher2,cantrell,hicks,zianni,fletcher3,lei}
as well as bulk materials \cite{lei,xing}.
Almost all of the earlier theoretical treatments were based on
the Boltzmann equation in analyzing the diffusion
component of thermoelectric power in macroscopic systems
\cite{conwell,abrikosov,askerov,kearney} and the
phonon-drag component of it \cite{bailyn,cantrell,hicks,zianni}.
In a recent letter \cite{lei} Lei pointed out that thermoelectric power
can be conveniently analysed entirely within
the framework of balance equation approach.
He discussed thermoelectric power of both bulk materials and
quantum wells in the presence of carrier heating with a strong applied
electric field using balance equations for weakly nonuniform
systems \cite{leixie}. His calculation indicates
that hot-electron effects on thermoelectric power may not only change
its magnitude, but also may change its sign at
higher electric field.  His result was confirmed very recently
by Xing, Liu, Dong, and Wang \cite{xing} using the
non-equilibrium statistical operator method of Zubarev \cite{zubarev}
jointly with the Lei-Ting balance equation approach \cite{leiting}.
However in assuming that the phonon-drag contribution to thermoelectric
power may be neglected in the range of electron temperatures of interest
for hot-electron transport \cite{xing}, both of their
treatments \cite{lei,xing} missed this contribution, which is
known to be very important in linear transport
in both bulk semiconductors \cite{geballe} and two-dimensional (2D)
systems \cite{nicholas,fletcher1,fletcher2,fletcher3}. It is
usually so large that the diffusion contribution would not be
discernible unless measurements were extended to very high or
very low temperatures \cite{fletcher3}.

In our opinion this assumption is questionable. The analysis reported
here includes the phonon-drag component $S_g$ of
thermoelectric power using the balance equation approach, which
has the advantage of easy inclusion of hot electron effects. In this,
we clarify the importance of the contribution of $S_g$ to
thermoelectric power in the hot-electron transport condition. Our
consideration are applicable to the regime in which the electron drift
velocity is lower than the sound velocities, for materials having high
impurity concerntratinos and/or intermediate electric field strength.
In fact, our result
shows that contrary to the assumption of Xing {\em et al.} \cite{xing},
$S_g$ is markedly enhanced at low lattice temperature under the
indicated conditions. As  the
calculations of Lei \cite{lei} and Xing {\em et al.} \cite{xing} both show
that the hot-electron effect does not change the absolute value of
the diffusion component of thermoelectric power very much, it is
therefore clearly important to include $S_g$ in the hot-electron
transport condition, provided that the lattice temperature is not too
high.

This paper is organized as follows. In Sec.\ II, we present
hydrodynamic balance equations with the phonon-drag effect included.
In Sec.\ III, we compare our theoretical prediction of $S_g$
with experiments in the linear transport limit.
Finally, in Sec.\ IV, we discuss hot-electron effects on $S_g$
and present our conclusions.

\section{Hydrodynamic balance equations including phonon-drag}

The balance equations for electron transport under the influence of an
electric field ${\bf E}$ and in the presence of a small lattice
temperature gradient $\nabla T$, can be written as \cite{leixie}:
\begin{eqnarray}
\label{diffusion}
&&\frac{\partial n}{\partial t}+\nabla\cdot({\bf v}n)=0\;,\\
\label{force}
&&\frac{\partial {\bf v}}{\partial t}+{\bf v}\cdot
\nabla{\bf v}=-\frac{2}{3}
\frac{\nabla u}{mn}+\frac{e{\bf E}}{m}+\frac{{\bf f}_i+
{\bf f}_{ep}}{mn}\;,\\
\label{energy}
&&\frac{\partial u}{\partial t}+{\bf v}\cdot\nabla u=-\frac{5}{3} u
(\nabla\cdot{\bf v})-w-{\bf v}\cdot({\bf f}_i+{\bf f}_{ep})\;.
\end{eqnarray}
In these equations the carrier drift velocity ${\bf v}$, the electron
temperature $T_e$, the average relative electron energy $u$, the carrier
density $n$ and the chemical potential $\mu$ are
all weakly dependent on spatial coordinates due to the small lattice
temperature gradient, so that their spatial gradients are all
small. Eqs.\ (\ref{diffusion})-(\ref{energy})
are accurate to the first order in these small quantities (all the
second and higher orders of spatial derivatives are neglected). The
frictional forces due to impurity scattering ${\bf f}_i$ and due to
phonon scattering ${\bf f}_{ep}$, and the energy transfer rate
from the electron system to the phonon system $w$, are all functions
of local quantities $n$, ${\bf
v}$, $T_e$ and $T$. ${\bf f}_i$ has exactly the same expression as in the
uniform case \cite{lei}. ${\bf f}_{ep}$ and $w$ can be written as ($\hbar
=k_B=1$)
\begin{eqnarray}
{\bf f}_{ep}&=&-4\pi\sum_{{\bf kq}\lambda} |\bar{M}({\bf q},\lambda)|^2
{\bf q}\,\delta (\varepsilon_{{\bf k}+{\bf q}}-\varepsilon_{\bf k}+
\Omega_{{\bf q}\lambda}-{\bf q}\cdot{\bf v})\nonumber\\
\label{fp}
&&\hspace{1cm}\mbox{}\times\{n_{{\bf q}\lambda}[f(\varepsilon_{\bf k},T_e)
-f(\varepsilon_{{\bf k}+{\bf q}},T_e)]-f(\varepsilon_{\bf k},T_e)[1-
f(\varepsilon_{{\bf k}+{\bf q}},T_e)]\}\;,\\
w&=&4\pi\sum_{{\bf kq}\lambda} |\bar{M}({\bf q},\lambda)|
^2 \Omega_{{\bf q}\lambda}\,\delta (\varepsilon_{{\bf k}+{\bf q}}-
\varepsilon_{\bf k}+\Omega_{{\bf q}\lambda}-{\bf q}\cdot{\bf v})
\nonumber\\
\label{w}
&&\hspace{1cm}\mbox{}\times\{n_{{\bf q}\lambda}[f(\varepsilon_{\bf k},T_e)
-f(\varepsilon_{{\bf k}+{\bf q}},T_e)]-f(\varepsilon_{\bf k},T_e)[1-
f(\varepsilon_{{\bf k}+{\bf q}},T_e)]\}\;,
\end{eqnarray}
where $\varepsilon_{\bf k}=k^2/2m$ is the energy dispersion of an electron
for an isotropic parabolic band system, $\Omega_{{\bf q}\lambda}$ denotes
the energy of a phonon of wavevector ${\bf q}$ and branch $\lambda$,
and $f(\varepsilon_{\bf k},T_e)=1/[\exp(\varepsilon_{\bf k}-\mu)/T_e+1]$
is the Fermi distribution function at the electron temperature.
$n_{{\bf q}\lambda}$ is the occupation number (distribution function) of
phonons in mode ${\bf q}\lambda$, which is weakly nonequilibrium due to the
existence of the small thermal gradient. It is
just this gradient that leads to a net flow of phonons in one
preferred direction, which then imparts momentum to electrons
via electron-phonon scattering. Therefore, in the balance equation
approach, the phonon-drag contribution to thermoelectric power is
embedded in the nonequilibrium phonon distribution function $n_{{\bf
q}\lambda}$ and thereby in the electron-phonon-scattering
frictional force ${\bf f}_{ep}$, which was previously taken to be
the equilibrium Bose distribution function $n(\Omega
_{{\bf q}\lambda}/T)=1/[\exp(\Omega_{{\bf q}\lambda}/T)-1]$ in
Refs.\ \onlinecite{xing} and \onlinecite{lei}.
$\bar{M}({\bf q},\lambda)$ stands for
the dynamically screened electron-phonon matrix element
\cite{leihoring,scalapino}. Considering the relation
\begin{eqnarray}
&&\delta (\varepsilon_{{\bf k}+{\bf q}}-\varepsilon_{\bf k}+
\Omega_{{\bf q}\lambda}-{\bf q}\cdot{\bf v}) f(\varepsilon_{\bf k},T_e)[1-
f(\varepsilon_{{\bf k}+{\bf q}},T_e)]\nonumber\\
&&\mbox{}\hspace{1.cm}
=\delta (\varepsilon_{{\bf k}+{\bf q}}-\varepsilon_{\bf k}+
\Omega_{{\bf q}\lambda}-{\bf q}\cdot{\bf v}) [f(\varepsilon_{\bf k},T_e)
-f(\varepsilon_{{\bf k}+{\bf q}},T_e)]\; n(\frac{\Omega_{{\bf q}\lambda}
-{\bf q}\cdot{\bf v}}{T_e})\;,
\end{eqnarray}
Eqs.\ (\ref{fp}) and (\ref{w}) can be rewritten as
\begin{eqnarray}
\label{fp1}
{\bf f}_{ep}&=&-2\sum_{{\bf q}\lambda}{\bf q}|\bar{M}({\bf q},\lambda)|^2
\Pi_2({\bf q},\Omega_{{\bf q}\lambda}-{\bf q}\cdot{\bf v}) [n_{{\bf q}
\lambda}-n(\frac{\Omega_{{\bf q}\lambda}-{\bf q}\cdot{\bf v}}{T_e})]\;,\\
\label{w1}
w&=&2\sum_{{\bf q}\lambda}\Omega_{{\bf q}\lambda}|\bar{M}({\bf q},
\lambda)|^2\Pi_2({\bf q},\Omega_{{\bf q}\lambda}-{\bf q}\cdot{\bf v})
[n_{{\bf q}\lambda}-n(\frac{\Omega_{{\bf q}\lambda}-{\bf q}
\cdot{\bf v}}{T_e})]\;,
\end{eqnarray}
in which $\Pi_2({\bf q},\Omega_{{\bf q}\lambda}-{\bf q}\cdot{\bf v})$
stands for the imaginary part of the density-density correlation function
\begin{equation}
\Pi({\bf q},\omega)=2\sum_{\bf k}\frac{f(\varepsilon_{{\bf k}+{\bf q}},T_e)
-f(\varepsilon_{\bf k},T_e)}{\varepsilon_{{\bf k}+{\bf q}}-
\varepsilon_{\bf k}+\omega+i0^+}\;.
\end{equation}

The phonon distribution function $n_{{\bf q}\lambda}$ for the nonuniform
system can be determined as follows. Its total rate of change
can be written as
\begin{equation}
\label{dndt}
\frac{d}{dt}n_{{\bf q}\lambda}=\left(\frac{\partial n_{{\bf q}\lambda}}
{\partial t}\right)_d+\left(\frac{\partial n_{{\bf q}\lambda}}{\partial t}
\right)_{ep}
+\left(\frac{\partial n_{{\bf q}\lambda}}{\partial t}\right)_{pp}\;,
\end{equation}
where the diffusion term is
\begin{equation}
\label{drift}
\left(\frac{\partial n_{{\bf q}\lambda}}{\partial t}\right)_d=-{\bf v}_p
({\bf q},\lambda)\cdot\nabla n_{{\bf q}\lambda}\;.
\end{equation}
To the lowest order in the thermal gradient, the distribution function
$n_{{\bf q}\lambda}$ on the right hand side of Eq.\ (\ref{drift}) can
be replaced by the equilibrium Bose distribution function:
\begin{equation}
\label{drift1}
\left(\frac{\partial n_{{\bf q}\lambda}}{\partial t}\right)_d^1=-{\bf v}_p
({\bf q},\lambda)\cdot\nabla n(\frac{\Omega_{{\bf q}\lambda}}{T})\;.
\end{equation}
In these equations the dependence of $n_{{\bf q}\lambda}$ on spatial
coordinate ${\bf r}$ has been suppressed and is understood;
${\bf v}_p({\bf q},\lambda)$
stands for the group velocity of a phonon in mode ${\bf q}\lambda$, and is
given by $\nabla_{\bf q}\Omega_{{\bf q}\lambda}$. The second term
on the right hand side of (\ref{dndt}) is the rate of change of $n_{{\bf q}
\lambda}$ due to electron-phonon interaction, which describes the absorption
and emission of phonons by electrons in transport. It has been proved that
this term reads \cite{leicui,leih}
\begin{equation}
\label{dnep}
\left(\frac{\partial n_{{\bf q}\lambda}}{\partial t}\right)_{ep}
=2|\bar{M}({\bf q},
\lambda)|^2\Pi_2 ({\bf q},\Omega_{{\bf q}\lambda}-{\bf q}\cdot{\bf v})
[n_{{\bf q}\lambda}-n(\frac{\Omega_{{\bf q}\lambda}
-{\bf q}\cdot{\bf v}}{T_e})]\;.
\end{equation}
This effect, together with the diffusion of phonons, tends to drive the
phonon system out of equilibrium. On the other hand, there is a trend to
drive these nonequilibrium phonons back towards equilibrium, which is
described by the last term in Eq.\ (\ref{dndt}):
\begin{equation}
\label{dnpp}
\left(\frac{\partial n_{{\bf q}\lambda}}{\partial t}\right)_{pp}=-\frac{1}
{\tau_p({\bf q},\lambda)}[n_{{\bf q}\lambda}-n(\frac{\Omega_{{\bf q}
\lambda}}{T})]\;.
\end{equation}
Here $\tau_p({\bf q},\lambda)$ is a wavevector and mode-dependent
relaxation time, which is assumed to be determined by (a) boundary
scattering
\begin{equation}
1/\tau_{B\lambda}=v_{s\lambda}/L\;,
\end{equation}
(where $v_{s\lambda}$ denotes the speed of sound in mode $\lambda$,
and $L$ is of the order of the macroscopic dimension of the specimen
\cite{casimir,callaway}), and by (b) phonon-phonon processes
\cite{herring,callaway}:
\begin{equation}
\label{taupp}
1/\tau_{pp}({\bf q},\lambda)=A_\lambda T^3\Omega_{{\bf q}\lambda}^2\;,
\end{equation}
whence
\begin{equation}
1/\tau_p({\bf q},\lambda)=1/\tau_{B\lambda}+1/\tau_{pp}({\bf q},\lambda)\;.
\end{equation}
In the steady state, the occupation number for each phonon mode is constant,
{\em ie.} $dn_{{\bf q}\lambda}/dt=0$, and
Eqs.\ (\ref{dndt})-(\ref{dnpp}) lead to
\begin{equation}
\label{nq}
n_{{\bf q}\lambda}=\frac{
\left(\frac{\partial n_{{\bf q}\lambda}}{\partial t}\right)^1_d
+\frac{1}{\tau_p({\bf q},\lambda)}n(\frac{\Omega_{{\bf q}\lambda}}{T})
+\frac{1}{\tau({\bf q}\lambda,\Omega_{{\bf q}\lambda}-{\bf q}\cdot{\bf v})}
n(\frac{\Omega_{{\bf q}\lambda}-{\bf q}\cdot{\bf v}}{T_e})}
{\frac{1}{\tau_p({\bf q},\lambda)}+\frac{1}{\tau({\bf q}\lambda,
\Omega_{{\bf q}\lambda}-{\bf q}\cdot{\bf v})}}\;,
\end{equation}
in which we have introduced a phonon-mode dependent
and electron-drift-velocity
dependent inverse electron-phonon scattering time
\begin{equation}
\label{tau}
\frac{1}{\tau({\bf q}\lambda,\Omega_{{\bf q}\lambda}-{\bf q}\cdot{\bf v})}
=-2|\bar{M}({\bf q},\lambda)|^2\Pi_2({\bf q},
\Omega_{{\bf q}\lambda}-{\bf q}\cdot{\bf v})\;,
\end{equation}
and the phonon drift component $\left(\frac{\partial
n_{{\bf q}\lambda}}{\partial t}\right)_d$ has been
replaced by Eq.\ (\ref{drift1}).
Substituting the distribution function (\ref{nq}) into Eqs.\ (\ref{fp1})
and (\ref{w1}), and taking the definition (\ref{tau}) into account, we have
\begin{equation}
{\bf f}_{ep}={\bf f}_g+{\bf f}_p\hspace{0.6cm},\hspace{0.6cm}w=w_p\;,
\end{equation}
where ${\bf f}_g$ is the frictional force due to phonon drag; ${\bf f}_p$
and $w_p$ are the frictional force and energy transfer rate due to the
phonon absorption and emission. They are defined by
\begin{eqnarray}
\label{fg}
{\bf f}_g&=&\sum_{{\bf q}\lambda}{\bf q}\left(\frac{\partial
n_{{\bf q}\lambda}}{\partial t}\right)^1_d\frac{\tau^{-1}
({\bf q}\lambda,\Omega_{{\bf q}\lambda}-{\bf q}\cdot{\bf v})}
{\tau^{-1}_p({\bf q},\lambda)+\tau^{-1}
({\bf q}\lambda,\Omega_{{\bf q}\lambda}-
{\bf q}\cdot{\bf v})}\;,\\
{\bf f}_p&=&\sum_{{\bf q}\lambda}{\bf q}\frac{1}{\tau_p({\bf q},
\lambda)+\tau({\bf q}\lambda,\Omega_{{\bf q}\lambda}-{\bf q}\cdot{\bf v})}
[n(\frac{\Omega_{{\bf q}\lambda}}{T})-n(\frac{\Omega_{{\bf q}\lambda}
-{\bf q}\cdot{\bf v}}{T_e})]\;,\\
w&=&w_p=-\sum_{{\bf q}\lambda}\Omega_{{\bf q}\lambda}
\frac{1}{\tau_p({\bf q},
\lambda)+\tau({\bf q}\lambda,\Omega_{{\bf q}\lambda}-{\bf q}\cdot{\bf v})}
[n(\frac{\Omega_{{\bf q}\lambda}}{T})-n(\frac{\Omega_{{\bf q}\lambda}
-{\bf q}\cdot{\bf v}}{T_e})]\;.
\end{eqnarray}

We consider a small lattice temperature gradient along the $x$ direction:
$\nabla T=(\nabla_x T,0,0)$. Then, in addition to the drift velocity
(current flow) and the applied external electric field along
the $y$ direction,
there will be a small drift velocity (current) and a small electric field
in the $x$ direction: ${\bf v}=(v_x,v_y,0)$ and ${\bf E}=(E_x,E_y,0)$.
Here, $v_x$ and $E_x$ are small and the spatial variations of all
quantities are assumed to be along the $x$ direction only. Following the
same scheme as Lei \cite{lei} by considering the particle number,
force and energy balance equations (\ref{diffusion})-(\ref{energy}) to
first order in the small quantities, we obtain, in the steady state
\begin{eqnarray}
\label{fx}
&&-\frac{2}{3mn}\nabla_x u+\frac{f_g^x}{mn}+\frac{eE_x}{m}+\frac{f_x}{mn}
=0\;,\\
\label{fy}
&&neE_y+f_y=0\;,\\
\label{wy}
&&w+v_y f_y=0\;,
\end{eqnarray}
which state the force balances in the $x$ and $y$ directions, together
with the energy balance, respectively, and ${\bf f}={\bf f}_i+{\bf f}_p$.
For small $v_x$, $f_x$ is proportional to it, and $\rho=-f_x/(n^2e^2v_x)$
is the resistivity in the $x$
direction in the presence of drift velocity $v_y$ in the $y$ direction.
Then Eq.\ (\ref{fx}) may be rewritten as
\begin{equation}
\label{jx}
j_x=\frac{E_x}{\rho}-\frac{2}{3}\frac{\nabla_x u}{ne\rho}+\frac{f_g^x}
{ne\rho}\;.
\end{equation}
Noting that the energy $u$ and number $n$ densities are given
separately by
\begin{equation}
u=2\sum_{\bf k}\varepsilon_{\bf k}f(\frac{\varepsilon_{\bf k}-\mu}
{T_e})\hspace{0.6cm},\hspace{0.6cm}n=2\sum_{\bf k}
f(\frac{\varepsilon_{\bf k}-\mu}{T_e})\;,
\end{equation}
we can rewrite Eq.\ (\ref{jx}) in the form
\begin{equation}
j_x=L^{11}(E_x-\nabla_x \mu/e)+L^{12}_d(-\nabla_x T_e)+L_g^{12}
(-\nabla_x T)\;,
\end{equation}
in which $L^{11}=1/\rho$. $L_d^{12}=\frac{1}{e\rho}[\frac{5F_{3/2}
(\zeta)}{3F_{1/2}(\zeta)}-\zeta]$ denotes the coefficient of the electron
diffusion contribution to the current with $\zeta=\mu/T_e$ and $F_\nu(y)
=\int_0^\infty x^\nu[\exp(x-y)+1]^{-1}dx$.
$L_g^{12}$ is the coefficient of the
phonon drag contribution which is obtained from Eq.\ (\ref{fg}) as
\begin{equation}
L_g^{12}=-\frac{1}{ne\rho}\sum_{{\bf q}\lambda}\frac{q_x^2}{q^2}
\frac{\Omega_{{\bf q}\lambda}^2}{T^2}\frac{\tau^{-1}
({\bf q}\lambda,\Omega_{{\bf q}\lambda}-\omega_0)}{\tau^{-1}_p
({\bf q},\lambda)+\tau^{-1}({\bf q}\lambda,\Omega_{{\bf q}\lambda}-
\omega_0)} n^\prime(\frac{\Omega_{{\bf q}\lambda}}{T})\;,
\end{equation}
where we have used the Debye-type spectrum for acoustic phonons:
$\Omega_{{\bf q}\lambda}=v_{s\lambda}q$ with $v_{s\lambda}$ being the
longitudinal sound speed in $\lambda$ mode and $\omega_0=q_yv_y$. The
thermoelectric power $S$ of hot electrons in the presence of a strong
current flow in the $y$ direction is then \cite{askerov,lei}
\begin{equation}
\label{s}
S=\frac{L_d^{12}}{L^{11}}(\frac{\delta T_e}{\delta T})
+\frac{L_g^{12}}{L^{11}}\equiv S_d+S_g\;.
\end{equation}
The diffusion component $S_d$ has already been obtained
by Lei \cite{lei} and Xing {\em et al.}
\cite{xing}. When there is no current flow in the $y$ direction ($\delta
T_e/\delta T=1$), it matches the result obtained from the
Boltzmann equation with a constant relaxation time assumption in
Ref. \onlinecite{hicks}. Furthermore,
\begin{equation}
\label{sg}
S_g=-\frac{1}{ne}\sum_{{\bf q}\lambda}\frac{q_x^2}{q^2}\frac{\Omega_{{\bf q}
\lambda}^2}{T^2}\frac{1}{1+\frac{\tau({\bf q}\lambda,\Omega_{{\bf q}\lambda}
-\omega_0)}{\tau_p({\bf q},\lambda)}} n^\prime(\frac
{\Omega_{{\bf q}\lambda}}{T})
\end{equation}
is  the phonon-drag component of thermoelectric power. Thus,
Eqs.\ (\ref{s})-(\ref{sg}) yield the thermoelectric power directly.

The hydrodynamic balance equations (\ref{diffusion})-(\ref{energy})
for weakly nonuniform systems are also applicable to 2D systems
(quasi-2D electrons coupled to
bulk phonons) if the prefactor $\frac{2}{3}$ in Eq.\ (\ref{force})
and $\frac{5}{3}$ in Eq.\ (\ref{energy}) are replaced
by unity and 2, respectively \cite{lei}. Moreover $|\bar{M}({\bf q},
\lambda)|^2$ in Eqs.\ (\ref{fp}) and (\ref{w}) is replaced by
$|\bar{M}({\bf q},\lambda)|^2|I(iq_z)|^2$. Here, $I(iq_z)$
stands for a form factor associated with confinement of
the sample in the $z$ direction\cite{leihoring}.
Specifically, for a quantum-well structure it reads
\begin{equation}
|I(iq_z)|^2=\pi^4\sin^2y/[y^2(y^2-\pi^2)^2]
\end{equation}
with $y\equiv q_za/2$ and $a$ is the width of the quantum well. Thus, the
diffusion component of thermoelectric power $S_d$ for a 2D
system takes the form
\begin{equation}
\label{sd2d}
S_d=\frac{1}{e}\left[\frac{2F_1(\zeta)}{F_0(\zeta)}-\zeta\right]
\frac{\delta T_e}{\delta T}\;.
\end{equation}
The phonon-drag component $S_g$ shares the same expression as that of
Eq.\ (\ref{sg}). We may put this equation in a more symmetrical form
when there is no current flow in $y$ direction, and correspondingly
$\omega_0=0$:
\begin{equation}
\label{sgwu}
S_g=-\frac{1}{2ne}\sum_{{\bf q}\lambda}\frac{q_\parallel^2}
{q^2}\frac{\Omega_{{\bf q}\lambda}^2}{T^2}
\frac{1}{1+\frac{\tau({\bf q}\lambda,\Omega_{{\bf q}\lambda})}
{\tau_p({\bf q},\lambda)}} n^\prime(\frac{\Omega_{{\bf q}
\lambda}}{T})\;,
\end{equation}
with ${\bf q}=({\bf q}_\parallel,q_z)=(q_x,q_y,q_z)$. This equation can
be shown to be equivalent to that derived from the Boltzmann equation
by Cantrell and Butcher \cite{cantrell} if ``1''
in the denominator of Eq.\ (\ref{sgwu}) is
omitted. This correspondence arises because Cantrell and Butcher use an
approximation ({\em ie.} see Eq.\ (41) in the first paper of
Ref.\ \onlinecite{cantrell}) which is analogous to neglecting the
electron-phonon scattering time
$1/\tau({\bf q}\lambda,\Omega_{{\bf q}\lambda}
-{\bf q}\cdot{\bf v})$ in our formula (\ref{nq}).
If we eliminate this approximation in their derivation, then
their final result may be proved to be equivalent to Eq.\ (\ref{sgwu}).
In our opinion,
this approximation is valid only at low temperature. If temperature is
high, this approximation yields a result that is too large.

\section{Comparison with experiments in the linear transport limit}

In this section we apply the present theory to GaAs/GaAlAs
quantum wells. Both
longitudinal and transverse acoustic phonon modes are coupled to
electrons via deformation potential and piezoelectric interactions.
Only the longitudinal mode gives rise to deformation potential
coupling, whereas both longitudinal and (two branches of) transverse modes
contribute to piezoelectric interaction. Their matrix elements are given
in Ref.\ \onlinecite{leihoring}. The GaAs/GaAlAs
material parameters used in our calculations are as follows:
volume density $d=5.31$\ g/cm$^3$, effective
mass $m=0.07m_e$, transverse sound velocity $v_{st}=2.48\times 10^3$\ m/s,
longitudinal sound velocity $v_{sl}=5.29\times 10^3$\ m/s, acoustic
deformation potential $\Xi=8.5$\ eV, piezoelectric constant $e_{14}=1.41
\times 10^9$\ V/m and low-frequency dielectric constant $\kappa=12.9$.
We note that some of the values of these parameters are still uncertain
({\em eg.} in the literature the values for $\Xi$ range from 7 to 16\ eV).
Nevertheless the above listed values have been used frequently in balance
equation theory in a variety of transport problems and the
results obtained are in good agreement with experiments \cite{leihoring}.
The coefficients $A_\lambda$ in Eq.\ (\ref{taupp}) can be obtained by
fitting to experimental data, and a particular choice will be identified
below.

In Fig.\ 1 we first fit Eqs.\ (\ref{s}), (\ref{sd2d}) and (\ref{sgwu})
to the experimental data of Sample 1 of
Fletcher {\em et al.} \cite{fletcher1}, where the electron sheet density
is $N=1.78\times 10^{15}$\ m$^{-2}$, and the length between the contacts
used to determine the thermoelectric power is $L=0.98$\ mm. As there
is no external electric field in the experiment,
$\delta T_e/\delta T=1$, $\omega_0=0$ and $T_e=T$.
In our fitting we take $A_l=1.68\times 10^{-21}$\ s\ K$^{-3}$ for
longitudinal phonons and $A_t=4.36\times 10^{-22}$\ s\ K$^{-3}$
for transverse modes in Eq.\ (\ref{taupp}). For comparison purposes, the
thermoelectric power components due to deformation potential coupling
$S_{gl}$, those due to longitudinal piezoelectric interaction
$S_{gpl}$ and those due to transverse piezoelectric interaction
$S_{gpt}$ are also plotted in the
same figure. We then apply the same parameters to the calculation of $S$
for the other two samples of
Fletcher {\em et al.} \cite{fletcher1}. The plot of $S$ {\em vs.} $T$
in Fig.\ 2 is for electron sheet density
$N=4.8\times 10^{15}$\ m$^{-2}$, with length between the
contacts $L=0.63$\ mm, and in Fig.\ 3 it is for
$N=6.82\times 10^{15}$\ m$^{-2}$, with
$L=2.98$\ mm. The widths of the quantum wells in Fig.\ 1-3 are all
100\ \AA. From the figures we can see that our theory is in good
qualitative agreement with the experimental data. Furthermore,
 our results in Fig.\ 1-3 show that, at low temperature, the two
branches of transverse phonons dominate the drag
effect through piezoelectric interaction with electrons, whereas the
longitudinal phonons contribute to $S_g$ mainly through the deformation
potential coupling. It is also seen that the
phonon-drag component dominates  thermoelectric power in the experiments.

We also apply our theory to a very recent experiment \cite{fletcher3},
where the electron density is $N=3.6\times 10^{16}$\ m$^{-2}$, which is
of an order of magnitude larger than that of the previous samples.
The width of the quantum well is $a=100$\ \AA  and $L=2.5$\ mm, as
determined in Ref.\ \onlinecite{fletcher3} using experimental data for
thermal conductivity. We use the {\em same material parameters and
coefficients} $A_\lambda$ as in
the previous cases. Figure 4 exhibits a comparison of
the experimental data with our theoretical result \cite{remark1}.
Again the theory fits well qualitatively with experiment. However,
this time the theoretical prediction is smaller than the
experimental data even at low temperature.

This experiment has also been explained by
Fletcher {\em et al.} in the same paper \cite{fletcher3} using
a formula derived from the Boltzmann equation by Cantrell and Butcher
\cite{cantrell}. The material parameters they used are chosen
to fit the experimental data at low temperature. However
these parameters fail to fit the earlier experimental results shown in
Figs.\ 1-3 at low temperature. In the calculation of
Fletcher {\em et al.} \cite{fletcher3}, the relaxation time
of phonons is treated as a constant and is determined from the thermal
conductivity measured in their experiment. Even though the
temperature in the experiment ranges from low temperature to very high
values, they continue to adopted the approximation of Cantrell and
Butcher \cite{cantrell} throughout, whereas we believe it
 to be valid {\em only at low temperature.}
In Fig.\ 5 we compare our theory with that of Fletcher {\em et al.},
wherein the chain curve is calculated from our theory and
the dotted curve is the theoretical prediction of Fletcher {\em et al.}.
It is seen that the peaks predicted by the two theories occur at almost
the same temperature. Nevertheless, the approximation used by Fletcher
{\em et al.} yields the height of
the peak of their theory to be larger than it should be, because this
approximation fails at high temperatures. This may be
understood from the fact that if we include  the same
approximation in our formula ({\em ie.} omitting ``1''
in the denominator of
Eq.\ (\ref{sgwu})), the result obtained ({\em ie.} solid curve)
predicts the same peak height found by Fletcher {\em et al.}.

\section{Hot-electron effects on the phonon-drag component $S_g$ and
conclusions}

This theory may be easily used to calculate
thermoelectric power in the presence of a strong external electric field.
In fact, although the diffusion component $S_d$ may be negative within a
small low-lattice-temperature range at higher electric field \cite{lei},
the phonon-drag component $S_g$ is still positive. Moreover, it should
be even more important to include the phonon-drag contribution to
thermoelectric power in the hot-electron transport regime.
Actually, $S_g$ is markedly enhanced at low lattice
temperature as shown in Fig.\ 6, for the sample
$N=1.78\times 10^{15}$\ m$^{-2}$ and in Fig.\ 7 for the much higher
electron density sample $N=3.6\times 10^{16}$\ m$^{-2}$. Our calculated
results shown in Fig.\ 6 and 7 assume high impurity density, such that the
electron drift velocity in hot electron transport is smaller than the
sound velocities ({\em ie}. we discard $\omega_0$ in Eq.\ (32)).
In Fig.\ 6, $S_g$ is plotted against lattice temperature $T$ as a solid
curve for hot-electron temperature $T_e=T+200$\ K,
as a dashed curve for $T_e=T+100$\ K and as
a chain curve for $T_e=T+10$\ K,
with all the other material parameters
being the same as those used in Fig.\ 1. For comparison, $S_g$
in the linear transport limit is also plotted in the same figure
(Fig.\ 6) as a dotted curve, with part of it
obtained in Fig.\ 1. From the figure one can see that $S_g$ is
markedly enhanced by hot-electron effects in the low lattice-temperature
regime (LLTR) with $T<20$\ K. However, in the high lattice-temperature
regime (HLTR) with $T>20$\ K,
$S_g$ is slowly reduced as $T_e$ increases. Another feature to be noted is
that for $T<6$\ K and $T>10$\ K, $S_g$, for $T_e=T+200$\ K
(solid curve), is a bit smaller than it is for $T_e=T+100$\ K
(dashed one). These properties can be well
understood from the inset of Fig.\ 6, where $S_g$ is plotted as a
function of $\Delta T=T_e-T$ for two fixed typical lattice temperatures
$T=4$ and 40\ K of the two regimes. It is seen that in LLTR,
the hot-electron effect first increases $S_g$ very
rapidly as $T_e$ rises from $T$ to about several hundred degrees Kelvin
(104\ K in the inset for $T=4$\ K) and then gradually reduces it if
$T_e$ goes on increasing. Nevertheless, even though $T_e$
is as high as 1004\ K as shown in the inset of Fig.\ 6, $S_g$ is still
so large at low lattice temperature that it can not be neglected as
suggested by Xing {\em et al.}\cite{xing}. However in HLTR,
$S_g$ decreases slowly as $T_e$ increases. The results for $S_g$ in regard
to hot electron effects are somewhat different at higher density for
the sample of Fig.\ 7 with $N=3.6\times 10^{16}$\,m$^{-2}$. In this
case, $S_g$ is
plotted versus $T$ as a solid curve for $T_e=T+100$\ K and as a
dashed one for $T_e=T+10$\ K, with all the other material
parameters being the same as those
used in Fig.\ 4. For comparison we also plot $S_g$ in the linear transport
condition as a dotted curve in the same figure, which has been obtained in
Fig.\ 4. Again $S_g$ is markedly enhanced in LLTR for $T<10$\ K.
However, for $T_e=T+100$\ K (solid curve), $S_g$ is now
very close to the plot for $T_e=T+10$\ K (dashed curve) in LLTR
and $S_g$ seems
to be almost unchanged in HLTR. This is further illustrated in the
inset of Fig.\ 7, where $S_g$ is plotted as a function of $\Delta T=T_e-T$
for three fixed lattice temperatures $T=3$, 6, and 15\ K.
It is seen that $S_g$ increases to ``saturation'' steeply in LLTR as $T_e$
increases and then rises very slowly as electron temperature goes up to
several hundred degrees Kelvin. As $T_e$ further increases, $S_g$
begins to decrease slowly. Notwithstanding the fact that $T_e$ is as
high as $T+1000$\ K and $S_g$ is slowly decreasing, the absolute value
of $S_g$ is still much higher than that in the linear-transport
limit ($\Delta T=0$). In HLTR, $S_g$ is almost unchanged as $T_e$ rises to
several hundred degrees Kelvin before it begins to decrease. These new
features result from the high density of the electron gas.

It is well known that the phonon-drag component of thermoelectric
power is very important in the linear transport limit and is
usually so large that the diffusion contribution would not be discernible
unless measurements were extended to very high or very low
temperatures \cite{fletcher3}. Both the theoretical
calculations of Lei \cite{lei} and Xing {\em et al.} \cite{xing} show
that the hot-electron effect does not change the absolute value of
the diffusion component of thermoelectric power significantly. However,
our results show that {\em it is necessary to include the
phonon-drag component $S_g$ under hot-electron transport
conditions, provided that the lattice temperature is not too high, and
the electron drift velocity is smaller than the sound velocities, to
accurately calculate the thermoelectric power}.

In summary, we have determined the role of phonon-drag in
thermoelectric power under hot electron transport conditions using
balance equation theory for weakly nonuniform systems.
Having introduced wavevector- and phonon-mode-dependent relaxation
times due to phonon-phonon interactions,
the formula obtained can be used not only at low lattice temperature
(where the phonon relaxation time is determined
by boundary scattering), but also
at high temperatures where phonon-phonon scatterings are dominant in
determining the phonon relaxation time.
Our results  compare well with experimental data in
the linear transport limit for a wide range of electron densities
and are in good correspondence with the earlier
theoretical prediction \cite{cantrell,fletcher3}
 of the Boltzmann equation, appropriately modified.
The role of hot-electron effects on the phonon-drag contribution to
thermoelectric power $S_g$ has been clarified, and the
importance of the contribution of $S_g$ to thermoelectric power
under hot-electron transport conditions was stressed. In fact, the sign of
the total thermoelectric power $S$ in the presence of a sufficiently
strong external electric field depends on
competition between the negative $S_d$ and positive $S_g$ values for
high impurity-density materials.
For such materials, at least for the low lattice temperatures,
the negative $S_d$ value may be entirely counteracted by $S_g$,
 so that the total thermoelectric power remains positive.

\acknowledgements

The author would like to thank Professor X.L. Lei for valuable discussions.
Ms. W. Sun is also acknowledged for doing part of the
numerical calculations. This research is supported partially by U.S. Office
Naval Research (Contract No. N66001-95-M-3472).

\centerline{\large FIGURES}
\bigskip
\bigskip

FIG.\ 1. Variation of thermoelectric power $S$ as a function of
temperature $T$. The black dots are the experimental data of Sample 1 in
Ref.\ \onlinecite{fletcher2}. The solid curve is the theoretical result
obtained from Eqs.\ (31), (35) and (34) with $A_\lambda$ in
Eq.\ (\ref{taupp}) chosen to fit the experiment. The components due to
transverse piezoelectric coupling $S_{gpt}$ (short-dashed curve),
longitudinal deformation potential coupling $S_{gl}$ (dotted curve)
and longitudinal piezoelectric interaction $S_{gpl}$ (long-dashed curve),
together with the diffusion component $S_d$ (chain line), are plotted in
the same  figure. The parameters we employ are listed in the text.
\bigskip

FIG.\ 2. Thermoelectric power $S$ vs. temperature $T$. The black dots
are the experimental data of Sample 2 in Ref.\ \onlinecite{fletcher2}. The
 solid curve is our theoretical prediction using the same
 values of $A_\lambda$
in the calculation as those obtained from Fig.\ 1. The components of $S$
are represented by the other curves in the same way as in Fig.\ 1.
\bigskip

FIG.\ 3. Thermoelectric power $S$ is plotted as a function of temperature
$T$. The black dots refer to the experimental data of Sample 3 in
Ref.\ \onlinecite{fletcher2}. The other curves represent
components of $S$ as in Fig.\ 1.
\bigskip

FIG.\ 4. Comparison of thermoelectric power as given by the theoretical
prediction with experimental data. The open circles are the recent
experimental
data of Fletcher {\em et al.}\cite{fletcher3} . The curves represent $S$
and its components as in Fig.\ 1.
\bigskip

FIG.\ 5. Comparison of the Boltzmann equation predictions of
Fletcher {\em et al.}\cite{fletcher3} (with the approximation explained
in the text) with our balance equation predictions for $S$.
In the figure the chain curve is the theoretical prediction of our
balance equation formulation;
the solid curve is obtained from our formulas (\ref{s}), (\ref{sgwu}) and
(\ref{sd2d}), however, with the approximation of Cantrell and
Butcher\cite{cantrell} installed as explained in the text, following
Fletcher {\em et al.};  and the dotted curve is taken
from Ref.\ \onlinecite{fletcher3}
as the theoretical prediction of Fletcher {\em et al.}
\bigskip

FIG.\ 6. Phonon-drag component $S_g$ vs. lattice temperature $T$ for
the sample $N=1.78\times 10^{15}$\ m$^{-2}$. Solid
curve: in the hot-electron transport condition $T_e=T+200$\ K; dashed
curve: $T_e=T+100$\ K; chain curve: $T_e=T+10$\ K;
and dotted curve: in the linear transport limit. Inset: $S_g$ is plotted
as a function of $\Delta T=T_e-T$ for two fixed lattice
temperatures $T=4$\ K and 40\ K. The material parameters
are the same as those used in Fig.\ 1, as described in the text.

\bigskip

FIG.\ 7. Phonon-drag component $S_g$ vs. lattice temperature $T$ for
the sample $N=3.6\times 10^{16}$\ m$^{-2}$. Solid
curve: in the hot-electron transport condition $T_e=T+100$\ K; dashed
curve: $T_e=T+10$\ K;
and dotted curve: in the linear transport limit. Inset: $S_g$ is plotted
as a function of $\Delta T=T_e-T$ for three fixed
lattice temperatures $T=3$\ K, 6\ K,
and 15\ K. The material parameters
are the same as those used in Fig.\ 4 as described in the text.

\begin{references}
\bibitem{remark1} In this figure the diffusion component $S_d$ is taken as
$0.41 T$\ $\mu$V/K. However from Eq.\ (\ref{sd2d}) one would find
$S_d=0.396 T$\ $\mu$V/K. This is a consequence of the fact that
balance equation theory yields the diffusion component of
thermoelectric power as being independent of
scattering, as explained in Lei's paper \cite{lei}.
\end{references}
\end{document}